# Water Absorption and High Electric Conductivity in ß-PbO$_2$ and Ag$_5$Pb$_2$O$_6$

D. Djurek[**,1], Z. Medunić[1], M. Paljević[2], and A. Tonejc[3]

[1] A. Volta Applied Ceramics (AVAC), Kesten brijeg 5. Remete, HR-10000 Zagreb, Croatia
[2] Ruđer Bošković Institute, P.O. Box 180, HR-10000 Zagreb, Croatia
[3] Physics Department, Faculty of Science, University of Zagreb, P.O.Box 331, HR-10000 Zagreb, Croatia

## Abstract

Water absorption in lead peroxide ß–PbO$_2$ and Byström-Evers compound Ag$_2$·[Ag$_3$Pb$_2$O$_6$] has been investigated by infrared spectrometry and electric conductance measurements. Hydrated Ag$_5$Pb$_2$O$_6$ exhibits, between room temperature and 690 K, electric conductivity, at least, order of magnitude higher than pure silver.

**Corresponding author: e-mail: danijel.djurek@inet.hr

*NOTE:* The wire samples resembling Figure 5 are available upon request.

**Introduction**

Lead peroxide ß-PbO$_2$ is a white colored substance (1) which crystallizes in a tetragonal form and space group P4/mnm (2) with unit cell dimensions a = 495,56 and c = 338,67 pm (3,4). Pb$^{4+}$ cations are coordinated by 6 O$^{2-}$ anions in two mutually orthogonal octahedral chains as it is shown in Figure 1. In fact, ß-PbO$_2$ is commonly known as a black powder containing small amounts of water and/or OH$^-$ groups. For instance, authors in ref (3 ) report 0.3 weight percents of H$_2$O in their sample.

Ag$_5$Pb$_2$O$_6$ compound was firstly prepared (5) in trigonal form (space group *P31m*) by A. Byström and L. Evers (BE), while M. Jansen and coworkers (6,7,8) reported the hexagonal lattice with unit cell dimensions a = 593,24 and c = 641,05 pm (6). Pb$^{4+}$ cations are octahedrally coordinated by 6 O$^{2-}$ (Figure 2a) and channels stretched along the hexagonal c–axis contain two Ag(1) cations per unit formula (Figure 2b). In order to differentiate between the two types of Ag cations it is more suitable to write the formula as Ag$_2$·[Ag$_3$Pb$_2$O$_6$].

**Experiment**

In these experiments we used Alfa Aesar (Johnson Matthey) Puratronic$^®$ 99,999 % (metallic basis) ß-PbO$_2$ which is brown colored. The most abundant residual ingredients are declared as As (10 ppm), Nb (3 ppm) and Zn (3 ppm). The presence of the water was estimated by the use of the Perkin-Elmer infrared (IR) spectrometer and absorption spectrum of the sample (as delivered ß–PbO$_2$ powder mixed with KBr) is shown in Figure 3a. An independent IR evaluation was performed on the pure KBr in order to check possible water condensed on the spectrometer windows. OH stretching absorption band is positioned at 3430 cm$^{-1}$ and $H_2O$ bending absorption band at 1635 cm$^{-1}$, the latter should be compared to the bending absorption band at 1595 cm$^{-1}$ of $H_2O$ vapor. The subsequently dried powder manifests a weakening of both cited absorption bands and appearance of an absorption band at 1385 cm$^{-1}$ (Figure 3b), which persists up to decomposition temperature of ß–$PbO_2$ to $Pb_3O_4$ at 723 K. A due attention was paid to this band and X-ray diffraction analysis excludes the possibility of $CO^{3-}$ absorption band coming from the traces of $PbCO_3$, $PbCO_3·PbCO_3$ and $PbCO_3·2PbO$ as well as carbonates coming from the other metallic ingredients in ß–$PbO_2$. In order to prove this statement an additional test was done by the decomposition of ß–$PbO_2$ to $Pb_3O_4$ at 723 K in 32 bar $CO_2$ atmosphere which prevents the decomposition of the possible residual metallic carbonates. The decomposition product exhibited no IR absorption at 1385 cm$^{-1}$. The feature at 1385 cm$^{-1}$ may be attributed to an absorption coming from the bending mode of $H_2O$ molecules, which possibly replace the oxygen atoms in ß-$PbO_2$. The possibility of such a doping was reviewed by A. Byström (9), while in the works of W. Mindt (10) and F. Lappe (11) the possibility of $OH^-$ substitution was pointed out, although the method reported in (10) could not differentiate between $H_2O$ molecules and $OH^-$ groups.

An exposure of ß–$PbO_2$ to a 0.01 bar of $H_2O$ vapor at 350 K for 10 days results in an increase of the relative intensity of 1386 cm$^{-1}$ absorption as compared to the stretching absorption at 3445 cm$^{-1}$ (Figure 3c) and the powder appeared to be black (wet ß–$PbO_2$).

The possible presence of water in $Ag_2·[Ag_3Pb_2O_6]$ was also pointed out in the original paper of Byström and Evers (5), and we have tested this possibility, as follows.

In the first step $Ag_2·[Ag_3 Pb_2O_6]$ was prepared from the dried Puratronic® ß–$PbO_2$ powder and $Ag_2O$ purissimus grade with no water present, even on the IR scale. Both powders were then mixed in the atomic proportion Ag/Pb = 5/2 in an agat mortar and the mixture dried in air at 200 °C, in order to remove the water involved during the mixing. The mixture was then fired for 10 days at 603 K and 270 bar $O_2$ (medical grade) pressure. Thermogravimetric (TG) decomposition to $Pb_3O_4$ performed at 723 K shown in Figure 4a confirms the oxygen content $O_6$ and X–ray diffraction angles are in agreement with those reported by Jansen and coworkers (6).

Wet $Ag_2·[Ag_3 Pb_2O_6]$ prepared from the wet ß–$PbO_2$ by above described method exhibits IR absorption spectrum as shown in the Figure 3d, and two types of the involved water may be recognized: (a) the water adsorbed on the grain boundaries exhibiting an IR absorption at 3391 $cm^{-1}$ (OH stretching band) and 1656 $cm^{-1}$ ($H_2O$ bending band), and (b) the dopant $H_2O$ positioned in Pb–O octahedral sites exhibiting the absorption band at 1385 $cm^{-1}$ and the possible OH stretching band recognized as a weak absorption band at 2931 $cm^{-1}$.

TG decomposition curve of the wet $Ag_2·[Ag_3Pb_2O_6]$ performed in air is shown in the Figure 4b where four decomposition steps are visible; (a) 450–500 K when adsorbed water is released, (b) 625–670 K when dopant water molecules leave their sites, (c) 720–760 K when BE phase decomposes to $Pb_3O_4$ and Ag. $Pb_3O_4$ is further decomposed to tetragonal PbO at 885 K (d). The IR absorption analysis (Figure 3e) performed on the powder of the BE phase heated up to temperature 693 K, just below the decomposition temperature 723 K, and subsequently cooled to room temperature (RT) reveals an absence of the absorption feature at 1385 $cm^{-1}$ visible before heating, while the X-ray diffraction data confirm the integrity of the original BE phase. Labels (x) and (y) denote absorption bands of $H_2O$ and $CO_2$ respectively in air.

The measurements of the electric resistance on the ß–$PbO_2$ and dry BE phase were performed on the pellets compacted under the 3 kbar of the uniaxial pressure and by the use of the four probe method with the silver paint contacts. The dried Puratronic ß–$PbO_2$ pellet exhibited the resistivity at RT higher than $10^2$ ohm·cm which sounds that pure ß–$PbO_2$ is probably insulator or semiconductor (12). The pellet prepared from the wet ß–$PbO_2$ exhibited resistivity at RT ~ $82·10^{-6}$ ohm·cm. The electric resistivity of the dried BE phase exceeds at RT $60·10^{-3}$ ohm·cm, which is rather high value and two orders of magnitude higher than that reported in Ref. 7.

The measurement of the electric resistance of the wet $Ag_2·[Ag_3Pb_2O_6]$ introduces complications since, as it was stressed in Ref 13, higher uniaxial stress converts the compound into high resistivity state, while an application of the shear strain to the pellet favors the low resistivity state (1). In order to do the manipulation easier we used copper tubes (13) filled with the powders of the wet BE phase, and subsequently extruded to small diameters, which results in a highly conducting wires exhibiting a long term stability and mechanical strength.

The temperature dependence of the resistance of copper tube filled with the wet $Ag_2 \cdot [Ag_3Pb_2O_6]$ (outer diameter ID = 0.2 mm, inner diameter ID = 0.065 mm, voltage contacts distance D = 68 mm and filling factor 0.18 of the theoretical density) is shown in Figure 5a. Voltage contacts were made by the use of silver paint and an empty copper tube of the same diameter was mounted, together with the filled one, as a reference sample. Measuring current was 1 A. At 690 K resistance (a) starts to increase, as a result of the loss of the $H_2O$ dopant. The cooling to RT results in a standard linear dependence of the copper resistance (b) with the calculated resistivity at 295 K ~ $1.74 \cdot 10^{-6}$ ohm·cm. It is noteworthy that the temperature coefficient of the copper resistivity dependence (a) is equal to that of the resistivity dependence (b) (0.0045 $deg^{-1}$), which sounds for the possibility that parts of the copper tube are short circuited by the highly conducting spots of the BE phase. The barely calculated resistivity of the BE core is ~ $0,13 \cdot 10^{-6}$ ohm·cm at RT, and accounting for a low filling factor, electric resistivity of the BE compound is roughly estimated to be less than ~ $3 \cdot 10^{-8}$ ohm·cm.

Both compounds ß–$PbO_2$ and $Ag_2 \cdot [Ag_3Pb_2O_6]$ absorb the considerable amounts of the water several days after drying and outgassing at 693 K. Adsorption proceeds even in properly closed vessels. Wire samples presented in Figure 6 and Figure 7 of Ref 13 improved the electric conductance in the course of time (0.3 – 0.5 promiles per day), though the ends of BE filled tubes are closed by welding. Such a temporal increase of the conductance may be the result of the accommodation of the shear strain after wire extrusion, but also a gradual move of the adsorbed water to octahedral sites. Both effects may also be combined.

**Discussion and conclusion**

A favorable assumption is that $H_2O$ molecule replaces the lateral oxygen in Pb–O octahedron with hydrogen atoms directed toward two apic oxygen atoms. In BE phase the replaced oxygen atoms may be those in the hexagonal channel (denoted by x in Figure 2b). Strongly asymmetric Lorentzian of OH stretching band of adsorbed water at 3391 cm$^{-1}$ (Figure 3d ) suggests the presence of the corresponding band at lower frequency coming from the dopant $H_2O$. It would be hard to accept that the feature at 1386 cm$^{-1}$ represents the counterpart to the bending band of the free water (1595 cm$^{-1}$) since it is known that absorption frequency of this band is higher in the interstitial water molecules. Alternatively $H_2O$ molecule, as a part of the octahedron structure may be a clue for a novel type of calculation and interactions among more atoms involving possible new vibration modes.

The presence of the $H_2O$ molecule in the channel corner of the Pb–O octahedron may considerably affect the Ag(1) chain and create new physical situations. The distance between two Ag(1) cations is short (304 pm) which is the result of the $d^{10}$– $d^{10}$ interactions (15) and of the binuclear (tandem) nature of the Ag(1)–Ag(1) bond (16). This tandem structure provides only one valence for two Ag(1) cations and enables the electric neutrality of the unit cell. 1/6 of the lateral oxygen bond shared with Ag(1) cation (17) would disappear by $H_2O$ doping, part of Ag(1) atoms removed from the hexagonal chain, which in turn makes the material highly conducting.

In conclusion: (a) ß–$PbO_2$ and BE compound $Ag_2 \cdot [Ag_3Pb_2O_6]$ exhibit high resistivity when pure and stoichiometrically correct, (b) the wetting by the water reduces the resistivity of the ß–$PbO_2$ for more than 8 orders of magnitude, (c) preparation of $Ag_5Pb_2O_6$ from the wet ß–$PbO_2$ and $Ag_2O$ results in a high conducting state with electric resistivity from 300 – 690 K appreciably less than that of the silver, (d) IR absorption spectra of the wet ß–$PbO_2$ and $Ag2 \cdot [Ag_3Pb_2O_6]$ reveal two band types of the confined water; $H_2O$ adsorbed on grain boundaries and dopant $H_2O$ which substitutes lateral O atoms in Pb-O octahedra, (e) dopant $H_2O$ in both ß–$PbO_2$ and $Ag_2 \cdot [Ag_3 Pb_2O_6]$ possibly exhibits a pronounced IR absorption feature at 1386 cm$^{-1}$ .

Methods of a precise control of doping by the $H_2O$ molecules are on the course of the development.

The authors are grateful to ELKA facility for the extrusion of BE phase filled copper tubes.

**Figure captions:**

**Figure 1.** The crystal unit cell of the ß–$PbO_2$ and positions of two mutually perpendicular Pb-O octahedra.

**Figure 2.** (a) the crystal unit cell of the BE compound $Ag_2 \cdot [Ag_3Pb_2O_6]$ and (b) c-axis projection of the crystal lattice. (x) denotes oxygen atom possibly replaced by $H_2O$ molecule.

**Figure 3.** IR spectra recorded in air; (a) ß–$PbO_2$ Puratronic® as delivered, (b) dried ß–$PbO_2$ Puratronic®, (c) wet ß–$PbO_2$ (explanation in text), (d) wet $Ag_2 \cdot [Ag_3Pb_2O_6]$ (explanation in text), (e) $Ag_2 \cdot [Ag_3Pb_2O_6]$ outgassed at 693 K.

**Figure 4.** TG decomposition in air of (a) dried $Ag_2 \cdot [Ag_3Pb_2O_6]$ and (b) wet $Ag_2 \cdot [Ag_3Pb_2O_6]$.

**Figure 5.** Temperature dependence of the electric resistance of the copper tube filled by wet $Ag_2 \cdot [Ag_3Pb_2O_6]$, (a) heating and (b) cooling.

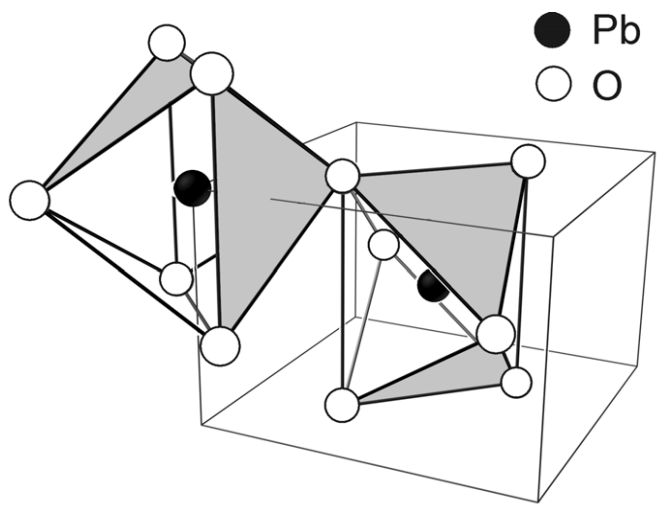

Figure 1

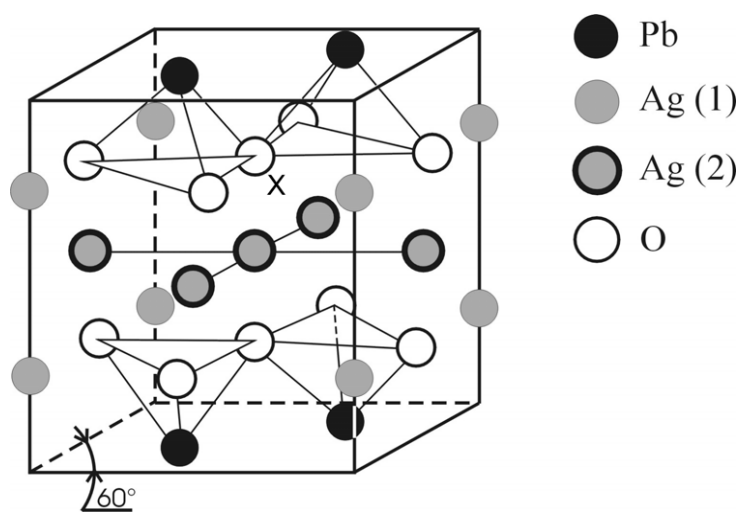

Figure 2a

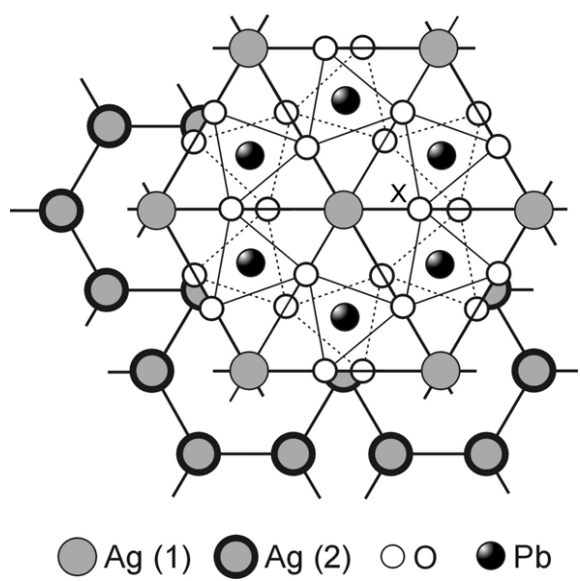

Figure 2b

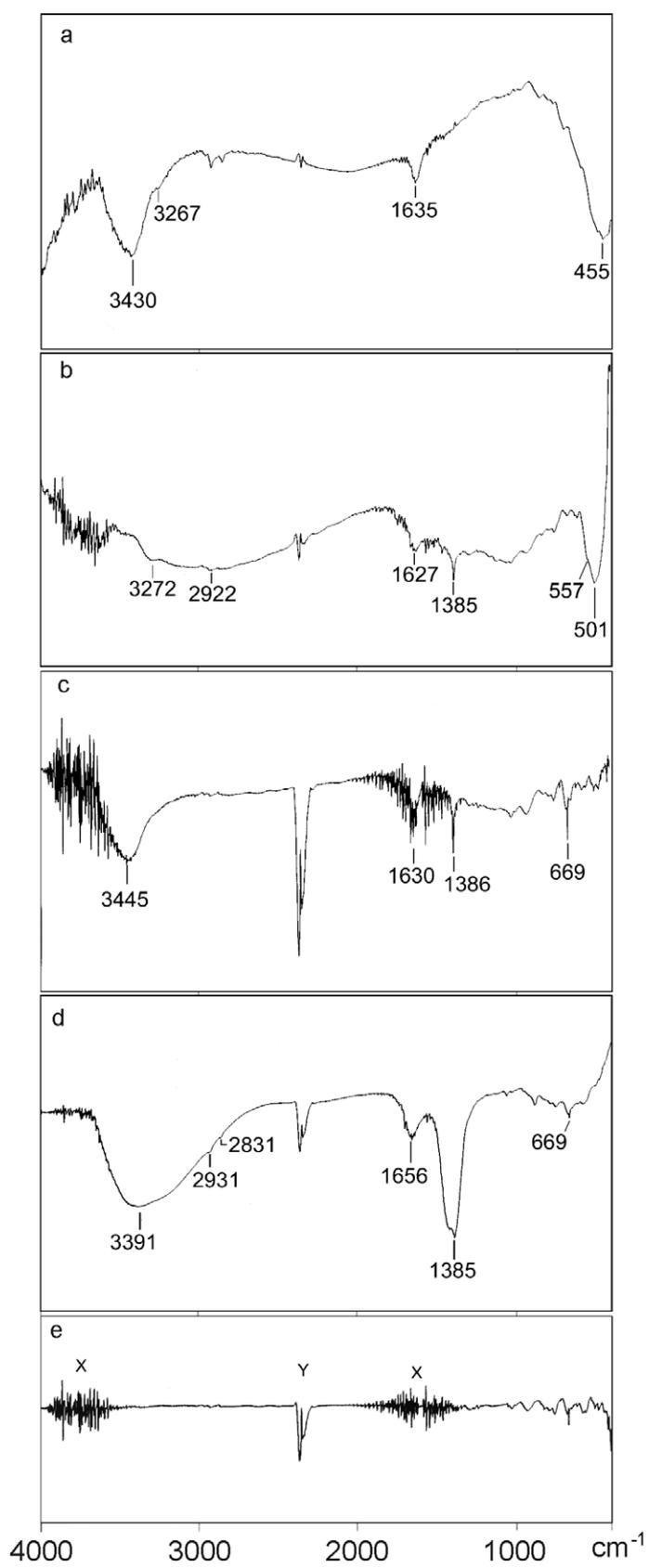

Figure 3

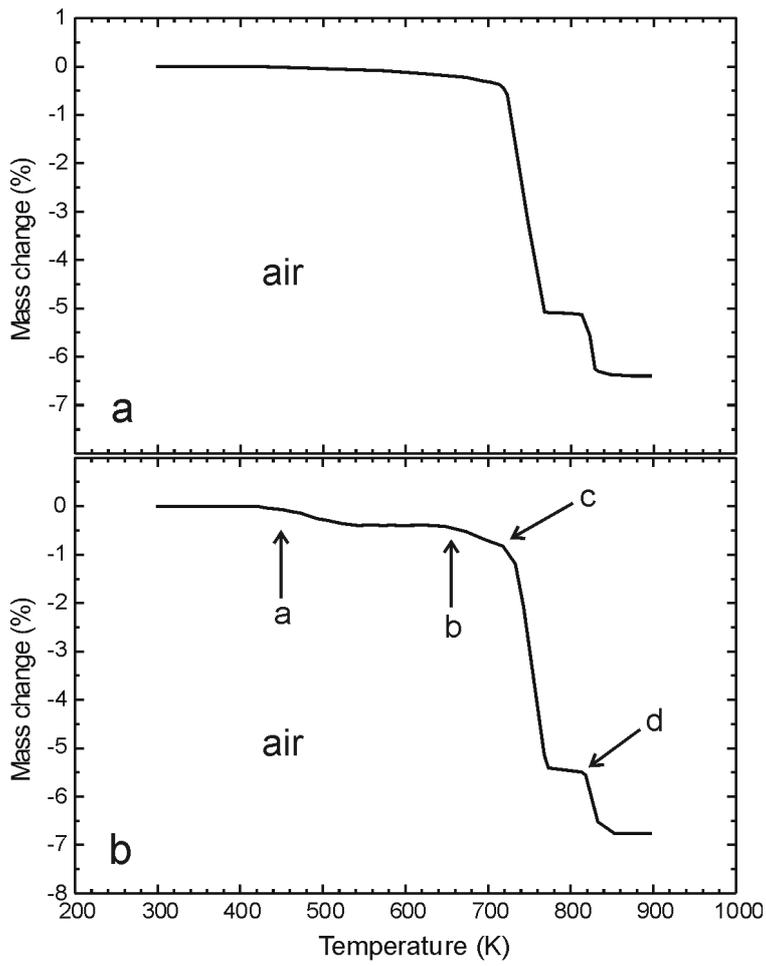

Figure 4

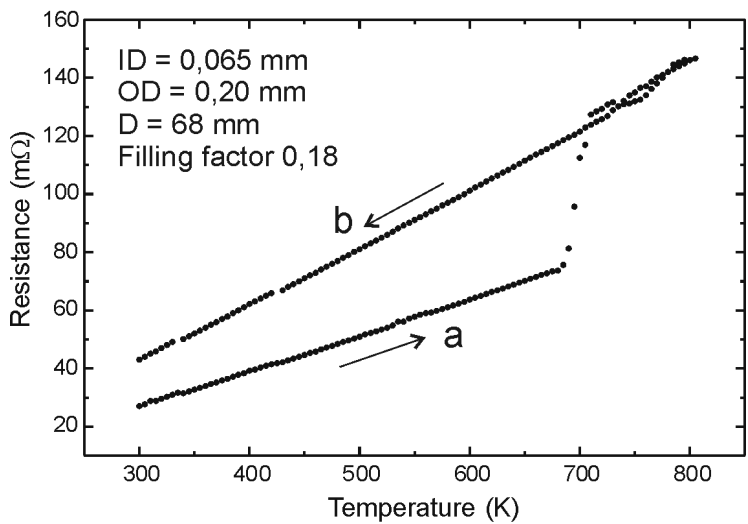

Figure 5